\documentclass{aastex}

\shorttitle{Pop III stars}
\shortauthors{Abia et al. }

\begin{document}

\title{The implications of the new Z=0 stellar models and yields \\
on the early metal pollution of the intergalactic medium}

\author{C. Abia\altaffilmark{1}, I. Dom\'\i nguez\altaffilmark{1}}
\affil{Dept. F\'\i sica Te\'orica y del Cosmos, Universidad de Granada,
    E-18071 Granada, Spain}
\email{cabia@ugr.es; inma@ugr.es}

\author{O. Straniero\altaffilmark{2}}
\affil{Osservatorio Astronomico di Collurania, I-64100 Teramo, Italy}
\email{straniero@astrte.te.astro.it}

\author{M. Limongi\altaffilmark{3}} 
\affil{Osservatorio Astronomico di Roma, Via Frascati 33, I-00040 Monteporzio Catone, Roma, Italy}
\email{marco@nemo.astro.mporzio.it}

\author{A. Chieffi\altaffilmark{4}}
\affil{Instituto di Astrofisica Spaziale (CNR), via Fosso Cavaliere, I-00133 Roma, Italy}
\email{achieffi@ias.rm.cnr.it}

\and

\author{J. Isern\altaffilmark{5}}
\affil{Institut d'Estudis Espacials de Catalunya - CSIC, Barcelona, Spain}
\email{isern@ieec.fcr.es}

\begin{abstract} 
Motivated by the recent detection of metals in different components of the high redshift 
universe and by the abundance ratios measured in the extremely metal-poor stars of our Galaxy, we study 
the nucleosynthesis constraints that this imposes on an early generation of stars (Population III).
To do so we take into account the chemical yields obtained from homogeneous evolutionary
calculations of zero metal stars in the mass range $3\la m/M_\odot\la 40$ (Limongi et al. 2000, Chieffi et al. 2001).
We also consider the role played by metal-free very massive objects (m$>100$ M$_\odot$).
Using both analytical and numerical chemical evolution models, we confront model
predictions from the different choices of the mass function  
proposed for Population III with the observational
constraints. We show that low values of star formation efficiency ($<1\%$) 
are required so as not to exceed the minimum metallicity ([C/H]$\approx -2.4$) measured
in the high redshift systems for any of the IMFs proposed. We also show that the 
observational constraints require $\Omega_{sr}< 3\times 10^{-3}\Omega_b$, confirming previous claims 
that the possible contribution of the stellar remnants from Population III to the baryonic dark 
matter is insignificant. At present, however, the scarcity of abundance measurements
for high redshift systems does not permit us to put severe limitations on the nature of the initial
mass function for Population III. In fact, overabundances of alpha-elements with respect to iron of the order 
of those measured in damped Lyman-$\alpha$ systems are obtained for any of the IMFs tested. Nevertheless, 
to account for the very large [C,N/Fe] ratios found in a considerable number of extremely metal-poor 
stars of our Galaxy, an IMF peaking at the intermediate stellar mass range ($4-8$ M$_\odot$) 
is needed.

\end{abstract}


\section{Introduction}

In current cold dark matter (CDM) hierarchic cosmological models, shortly
after the Big Bang (redshifts $z\sim 10-30$) the primordial gas cooled
rapidly and fragmented. From the first structures, with typical masses between 
$10^5$ and $10^8$ M$_\odot$, it is believed that a very early generation of
stars, Population III, formed. Since the chemical content of these
first structures was primordial (H, $^4$He and traces
of D, $^3$He and $^7$Li), this first generation of stars must have been formed essentially 
free of metals ($Z\sim 10^{-9}$).

The possible existence of Population III was originally postulated because of the presence of 
metal overabundances in Population II stars and especially due to the so-called
G-dwarf problem: the discrepancy between the number of observed low mass metal-poor
stars and that predicted by simple (closed) models of chemical evolution (Pagel 1997). This led to the idea 
that very shortly after the Big Bang and some time before the formation of the first 
galaxies ($z\sim 3$), a generation of stars polluted the universe with metals \citep{cay86}. 
Recent metal-poor stellar count surveys in the halo and thick disk of the Galaxy \citep{bee85, bee92} 
support this scenario, finding only a handful of stars with [Fe/H]$\leq -3$\footnote{In this work we use the
standard notation [X/Y]$\equiv$log (X/Y)$_{\rm{object}}-$log(X/Y)$_\odot$ for any
chemical species}. The observational evidence of a very low number 
of metal-poor low-mass stars suggests that the very first star formation episode
was biased towards large stellar masses. Extremely low-mass metal-poor stars are not seen simply because 
they did not form, i.e. if Population III were made up of very massive rapidly evolving stars, there
would be no possibility of observing them today. 

The existence of Population III is related to many important 
issues. This population of stars could play an
important role in the reionization of the universe \citep{mir98, gne00, fan00} compensating for
the fact that the known populations of quasars and star-forming galaxies are not
sufficient to account for the required number of ionizing photons at $z\geq 5$.
Alternatively, massive remnants (black holes) as the end-products of these primordial stars
could have been the progenitors of the current active galactic nuclei, while the less massive
remnants (ancient white dwarfs) could have made a contribution to the MACHO population found in the 
halo of our galaxy \citep{men00, alc00, las00} although this contribution is strongly limited
by the observed white-dwarf luminosity function as well as by energetic arguments \citep{ise98, ise99}. Obviously, 
the real significance of Population III stars in the above issues is determined by the 
nature of the initial mass function (IMF) of the first generation of stars.

The idea of a pregalactic metal enrichment has become increasingly important  in 
recent years because of the discovery of intergalactic metals in Ly-$\alpha$ forest 
clouds ($N(H I)\leq 10^{17}$ cm$^{-2}$) at $z= 3 - 3.5$ showing metal abundances of 
$Z\sim 10^{-3}-10^{-2} Z_\odot$ \citep{son97, cow98, ell00}.
\citet{sar80} were the first to identify the Ly-$\alpha$ forest with a population of primordial
hydrogen clouds in pressure equilibrium with a hotter, expanding intergalactic
medium (IGM). This picture, however, has changed recently to one in which the Ly-$\alpha$ forest
is, in fact, the IGM at very early epochs. Therefore, these high redshifted clouds
are believed to be essentially composed of unprocessed material from which
the galaxies were formed. However, the gas producing the Ly-$\alpha$ forest is not
pristine since a significant fraction of the detected lines are associated
with C IV absorptions at $\lambda\lambda 1548, 1550$ {\AA} \citep{tyl95, cow95, son96}. 
The carbon abundance implied by these detections is [C/H]$\sim -2.5$, although with an important
dispersion within the forest (see e.g. Borksemberg et al. 1998). Whether the metals
found have a local origin associated with the Ly-$\alpha$ lines observed, or are due to
an early episode of star formation (Population III) is still unclear. 
The answer to this question might come from metal abundance
studies in very low density clouds (N(H I)$\leq 10^{14}$ cm$^{-2}$). However, different
investigations have reached conflicting conclusions. \citet{lu98} concluded that
there is a sudden decrease of the C abundance at low column densities ([C/H]$\leq -3.5$ 
for N(H I)$< 10^{14}$ cm$^{-2}$). \citet{cow98}, however, claimed a roughly
constant C IV abundance distribution all the way down to Ly$-\alpha$ optical
depths $\tau(\rm{Ly}-\alpha)\sim 0.5-1$. More recently, \citet{ell00} have shown that the 
differences in redshift and velocity dispersion between Ly-$\alpha$ and C IV absorptions
prevent us, at present, from answering of whether there is a uniform degree
of metal enrichment down to the lowest values of N(H I).

An indirect way to infer the possible existence of Population III and the nature of
its stellar mass function is to search for the imprints that could have been left in the IGM and/or in
the subsequent generation of stars formed from the material ejected, that
is: the {\it chemical} imprint. In this paper we reexamine the issue of the chemical pollution 
that could have originated Population III. To do so, we make use of strictly 
zero metal ($Z=0$) stellar models in the mass range $3\leq m/M_\odot\leq 200$ (see below).
As an output, these models predict stellar yields for a number of significant isotopes 
which are the result of the nucleosynthetic processes 
occurring during the evolution of the stars. These processes include those occurring in the thermal-pulsing 
AGB phase (TP-AGB) for the intermediate stellar mass range ($3\la m/M_\odot\la 8$), in type II supernova 
(SNII) explosions ($m\ga 8$ M$_\odot$) and in very massive objects (VMO, $m\ga 100$ M$_\odot$).
 
The paper is structured as follows: The observational constraints used to limit our model predictions 
are discussed in section 2. The stellar yields for zero-metal stars are presented in section 3. Section 4 
describes the IMF proposed for Population III. In section 5, we use these tools to study the chemical 
consequences from a burst of Population III stars 
and from a simple chemical evolution model. Our conclusions are summarized in section 6.

\section{The observational constraints}

The main observational constraint concerns the detection of metals in the
Lyman-$\alpha$ forest spectra of distant QSOs. Since the Ly-$\alpha$ forest is being interpreted as intergalactic clouds, 
containing slightly contaminated primordial material, there must be some mechanism of mildly polluting them. Whether an 
early episode of pregalactic star formation (Population III)\footnote{We consider the Population III objects 
as {\it stars} with $Z=0$.} is required or not to explain these abundances is still unclear. As mentioned in $\S 1$, 
the key factor might be whether C IV lines continue to be seen in the Ly-$\alpha$ clouds of decreasing H I column density. 
Note, that a considerable fraction of the metals detected in the Ly-$\alpha$ forest systems might have
been produced by non zero-metal stars in already formed galaxies, and eventually  ejected into the IGM by
galactic winds. Here, we will assume as a {\it working hypothesis} that the metal contamination observed in the
Ly-$\alpha$ forest clouds is due to a pregalactic population of stars, i.e. Population III stars. Observations by 
\citet{cow98}, \citet{son97} and \citet{ell00} in the direction of highly redshifted QSOs have revealed line absorptions 
due to C, N, O and Si. Although the abundances derived are sensitive to ionization corrections \citep{sc00}, these studies 
obtained [C/H]$\sim -2.5$ and [Si/H]$\sim -2.3$ with a scatter of about a factor of 3 in systems with $z=3-3.5$ 
(no abundance ratios are given for N and O). Scaling these values to the solar metallicity implies that these systems 
have a metallicity of Z$\sim 4\times 10^{-3}$ Z$_\odot$, or [Fe/H]$\sim -2.4$. Note, however, that current observational 
techniques used to derive metal abundances in the Ly-$\alpha$ systems cannot properly account for the inhomogeneity of the IGM 
and, therefore, these techniques probably give only lower limits on Z. Nevertheless, as mentioned above, here we adopt 
these abundance ratios as the typical level of chemical enrichment in the IGM produced by the hypothetical Population III.

Our second observational constraint comes from the metal absorptions detected in very distant
systems with a high density of H I, N(H I)$\ga 2\times 10^{20}$ cm$^{-2}$, i.e.: the damped-Lyman systems.
In modern cosmological theories, damped systems are assumed to be more highly evolved
objects than the Ly-$\alpha$ forest and probably represent the earliest stages of the
present-day disc galaxies \citep{wol95}. These systems provide the best opportunity to measure 
abundances of different elements at high redshifts. Numerous abundance studies in these
systems [see Pettini (1999) for a compilation] have clearly established that the cosmic metallicity
inferred in neutral gas does not evolve significantly from $z\sim 1$ to 4 although the
unweighted metallicity exhibits a statistically significant decrease with increasing redshift
(Prochaska, Gawrser \& Wolfe 2001). The mean metal abundance in these systems between 
$z=1-4$ is $Z\sim 6\times 10^{-2}$ Z$_\odot$ (or [Fe/H]$\sim -1.2$),
although with a range of almost two orders of magnitude in the metallicity attained by different
damped systems at essentially the same redshift. The main difficulty encountered in these abundance 
studies is to account for the fraction of each element that has been removed from the gas phase to form
interstellar dust. Despite this, most of the abundance ratios measured are similar to the
abundance pattern expected from solely SNII pollution \citep{lu98}, although not all 
the abundance ratios derived follow this SNII pattern \citep{pro99, pet99}. The
limitations introduced by these abundance ratios upon the enrichment from 
Population III are not clear, since if they are considered more evolved objects, the abundance
pattern in damped systems may be the result of several episodes of star formation. Therefore, 
it does not reflect the result of the very first metal pollution.
We note, however, that no damped system is observed with [Fe/H]$<-2.7$ \citep{pro00}.

Finally, the abundance ratios found in the most extremely metal-poor stars
in our galaxy constitute another observational constraint
to the early chemical evolution. For instance, one of the most striking results of these studies is  
that $\sim 25\%$ of the stars with [Fe/H]$\la -2.5$ exhibit large carbon
and nitrogen abundances relative to iron, [C,N/Fe]$\ga 1$ \citep{nor00, ros99}.
Many of these metal-poor carbon/nitrogen enhanced stars also exhibit extremely high abundances
of neutron-capture elements \citep{sne96, bon98, dep00, sne00, hil00, spi00}. 
Despite the fact that their evolutionary status is not completely known, many of them 
are certainly unevolved (or turn-off) stars for which 
the hypothesis of internal pollution can be safely discarded (note that many of these
stars present Li abundances similar to those of the Spite plateau). One might
think that the entire CN-enhancement observed in these Population II stars is the result
of the mass transfer across a binary system which formerly contained an AGB star 
(in fact, many of them present radial velocity variations); there are, however, several
reasons for believing that this might not always be the case. The fact that $\sim 25\%$ of the
most metal-deficient stars are CN-enhanced comprises a stringent requirement on the fraction
of such systems that form binaries with the right configuration for the mass transfer to 
occur. Furthermore, not all of these stars present abundance patterns with the
s-process elements thought to occur in AGB companions. The abundance patterns
in these stars might well be content the key to understanding
the nucleosynthetic processes in the very early Galaxy before these extremely metal-deficient
stars formed.

\section{The IMF for Population III stars}

The issue of the IMF for the first generation of stars has received particular
attention (see e.g. Matsuda, Sato \& Takeda 1969; Carberlg 1981; Silk 1983). Since metals
are the most important coolants in present-day star formation, these studies have emphasized 
the importance of the radiative cooling by H-based molecules because the primordial 
gas was deficient in heavy elements. Current models predict that the first bound systems
capable of forming stars appeared in the history of the universe at redshifts between 50 and 10 and 
had masses between 10$^5$ and 10$^8$ M$_\odot$ (see e.g Miralda-Escud\'e 2000). 
Depending mainly on the ratio between the time scale for radiative cooling and heating due to the
gravitational collapse, these structures could further fragment into clumps with 
smaller masses. These simulations show that typical fragmentation into clumps occurs
with masses in the range 10$^2$-10$^3$ M$_\odot$, but due to the relatively low
resolution used in these studies, there is no clear consensus on whether these
structures can fragment into still smaller mass clumps. This controversy seems to
originate from oversimplified assumptions like a homogeneous, pressure-less
and/or spherical collapse. The minimum mass fragment that could form is of special relevance
because of the feedback effects. If the typical protostar fragment formed had a large mass
($\sim 10^2$ M$_\odot$), its high luminosity would contribute to the rapid heating
of the collapsing clump, inhibiting further fragmentation. Additionally, the final fate
of these high mass objects would probably be a very energetic phenomenon (SNII-like or
hypernova explosion). The large energy deposited in the gas during this event could
be large enough to completely disrupt the clump, again avoiding further star formation. Thus, the
typical mass of the collapsed fragment not only determines the IMF but also the 
period of time during which the first star formation episode occurred. 

In this respect, \citet{uea96} set a minimum fragment mass at about the Chandrasekhar mass 
($> 1.4$ M$_\odot$). This is an important result because it implies that we should see no
metal-free stars at the present time, since all of these stars should by now
have evolved. \citet{yos86}, based on the opacity-limited fragmentation theory,
derive an IMF that is steeper in the high stellar mass range than a Salpeter-like IMF;
the exact slope, however, critically depends on the mass-luminosity relation assumed for 
the zero-metal protostars. An interesting result from the Yoshii \& Saio calculations is
that the IMF derived would have a maximum around the intermediate stellar mass
range (3-8 M$_\odot$). Very recently,
\citet{nak00} performed multi-dimensional hydrodynamic simulations of the collapse
and fragmentation of filamentary primordial clouds. These simulations show that, depending upon the initial
density of the cloud, the IMF for Population III stars is likely to be bimodal. Gas filaments with initial
densities lower than $\sim 10^5$ cm$^{-3}$ tend to fragment into structures with masses larger than
several tens of $M_\odot$ ($\sim 10^2$ M$_\odot$), while initially denser filaments
($n\ga 10^5$ cm$^{-3}$) experiment more effective H$_2$ cooling and fragment
into structures of $\sim 1-2$ M$_\odot$. The relative peaks of this bimodal IMF would  
be a function of the collapse epoch in such a way that
during the first epoch the dominant peak would be around $2$ M$_\odot$, while
as the star formation proceeds and the collapsing clump heats up, the Jeans mass
would be displaced to larger masses, moving the peak to $\sim 10^2$ M$_\odot$.

The above mentioned studies basically compile the proposed IMFs for Population III stars
in the literature. In summary: i) there are physical arguments for assuming that the slope of the present
IMF has changed in time, ii) the first star-forming clouds typically fragmented into massive clumps of
$10^2-10^3$ M$_\odot$, iii) there exists the possibility of further
fragmentation into smaller masses, producing an IMF peaking at about the intermediate
stellar mass range. Support to these facts hasbeen recently given by 
Hernandez \& Ferrara (2001). In the framework of the standard hierarchical clustering scenario
of galaxy formation, these authors show that the IMF of the first stars was increasingly high mass
biassed towards high redshifts: at $z\sim 9$ the charactheristic stellar mass being 10-15 M$_\odot$.
Therefore, we will restrict our analysis to the IMFs that cover the above properties. In fact, we use the 
IMFs proposed by Yoshii \& Saio, referring to them as YSa if a relation mass-luminosity in the form 
$L(m)\propto (m/M_\odot)^\beta$ with $\beta=1.5$ is adopted, or YSb if $\beta=3$. We also 
use the bimodal IMF proposed by Nakamura \& Umemura and, in a similar way, 
we consider two cases, depending on the position of the peaks: NUa if the peak
at low stellar mass is the dominant one or NUb if the peak at large mass dominates.
For comparison, we contrast the results with those
obtained from the canonical \citet{sal55} IMF. Figure 1 shows the IMFs used in the
mass range 1-10$^3$ M$_\odot$, and clearly demonstrates how these IMFs differ 
\footnote{The IMFs are normalized as $\int_{m_{low}}^{m_{up}}\phi(m) dm\equiv 1$, 
where $m_{up}$ and $m_{low}$ are the upper and the lower stellar mass limit, respectively.
Note that the upper limit varies depending
upon the IMF adopted. From the \citet{uea96} results we have adopted $m_{low}=1$ M$_\odot$
in all the IMFs.}.

\section{The stellar yields}

Recent theoretical analysis of the evolution of Population III stars
predicts that the fate of metal-free stars can be classified as follows:
(1) Stars with initial mass $m\ga 250$ M$_\odot$ collapse to
a black hole (BH) with no metal ejection  \citep{obe83, heg00}.
(2) Stars in the mass range $100\la m/M_\odot\la 250$, usually
termed very massive objects (VMO), are disrupted by electron-positron
pair instability leading to a type II supernova-like event (pair creation supernovae, PCSN) and eject
metals leaving no compact remnant behind. (3) Stars with initial mass in 
the range $40\la m/M_\odot\la 100$ probably
collapse into a BH without metal contribution (Woosley \& Waever 1995, but see below). 
(4) Stars in the range $10\la m/M_\odot\la 40$ give a type II supernova (SNII) 
event \citep{wos95, lim01, lim00} and finally. (5) Stars with initial mass $m\ga 1$ M$_\odot$ 
and lower than the minimum mass for carbon ignition ($\sim 8$ M$_\odot$) will pass 
through the AGB phase \citep{fuj00, chi01} leaving a white dwarf (WD) as 
the stellar remnant.

The nucleosynthesis products from intermediate mass stars (IMS) ($1.5\la m/M_\odot\la 8$)
are of special interest here since the IMF for Population III stars
might peak in this mass range. Note that a star with initial mass around 7-8 M$_\odot$
has a lifetime ($\sim 10^7$ yr) only slightly larger than that of a typical SNII progenitor and
therefore will contribute to IGM enrichment shortly afterwards.
Due to the peculiarity of the $Z=0$ stellar evolution, appropriate yields are highly
necessary. Previous calculations of the stellar yields 
in metal-poor IMS were obtained from stellar models with no primordial composition, 
$Z\sim 10^{-4} Z_\odot$ \citep{mar98, van97}. Moreover, these yields were obtained
from synthetic approximations after the end of He-burning without having taken into account the
evolution during the TP-AGB phase. In fact, the existence of third dredge-up (TDU) episodes 
and thermal pulses (TP) in this phase plays a crucial role in determining the stellar yields from IMS. 
Recently, \citet{chi01} performed a detailed analysis of the evolution of zero metal IMS. Using the
evolutionary code FRANEC \citep{chi98} these authors found, at variance with previous
studies, that these stars do experience TP during the AGB phase. As a consequence,
TDU episodes occur and the stellar envelope may be enriched with fresh
$^{12}$C, $^{14}$N and $^{16}$O during this phase. The most important result from this study is 
that these stars become C-rich (i.e. carbon stars) and N-rich and can eject important amounts
of these elements at the end of their evolution. The yields from these stars are, however, sensitive
to the modeling of the mass-loss rate during the AGB phase. In fact, it is the mass-loss
rate that finally determines the duration of the AGB phase since it limits the number of TPs
and, therefore, the number of TDU episodes. Both mass loss and dredge up efficiency
could be deduced by the observed properties of the disk AGB population, but 
little is known of these phenomena in very metal-poor stars. We note, however,  
that the increase of the heavy elements in the envelope, caused by the TDU in Population III AGB stars, 
could reduce the difference with respect to the more metal-rich AGB stars. The yields of intermediate 
mass stars shown in Figure 2 have been obtained by assuming a Reimers mass loss rate 
with $\eta=4$. This is considered a reasonable average mass loss rate for Population I intermediate 
mass AGB stars (see van den Hoeck \& Gronewegen 1997). We emphasize that the $^{14}$N production from 
IMS is quite different from the value adopted here if a different value of $\eta$ is assumed.
For example, assuming $\eta\approx 1$ the $^{14}$N yield decreases by a factor $\sim 3$.   
On the other hand, Chieffi et al. (2001) also show that in the most massive models 
($m> 4$ M$_\odot$), $^7$Li can also be synthesized via hot bottom burning by the \citet{cam71} mechanism
in the same way as occurs in higher metallicity models \citep{sac92}. The Li mass
fraction in the envelope can reach peak values $\sim 10^{-8}$, although the final Li yield
would be lower than this value because of the progressive $^3$He consumption and Li
depletion at the base of the envelope during the AGB phase. 

Concerning the lowest mass range among IMS ($1.5\la m/M_\odot\la 3$), \citet{fuj00}
have shown that these stars develop TP and also become C and N-rich. Unfortunately, these
authors did not compute yields from this mass range. However, we believe they can be omitted from our analysis,
for the following reasons: i) Most studies of the fragmentation of primordial gas agree
that objects with mass $m< 2$ M$_\odot$ are very difficult to form. ii) We are mainly
interested in the very early chemical enrichment of the IGM. Stars with $m< 3$ M$_\odot$ have a 
long lifetime and, if they formed, would only contribute at very late times in the 
chemical evolution of the IGM.

The adopted stellar yields for the mass range $10\la m/M_\odot\la 40$ 
are taken from \citet{lim00}. These yields are based on presupernova evolutions computed
with the FRANEC code plus a simulated explosion based on the radiation dominated shock
approximation \citep{wos80}. The location of the mass cut has been chosen by requiring
that $\sim 0.05$ M$_\odot$ of $^{56}$Ni are ejected by each stellar model. This amount of Ni
is very similar to that used to explain the observational properties of the SN 1987A. Let us
remind that since the location of the mass cut is still theoretically very uncertain, the yields
of the elements mainly produced at the bottom of the mantle will largely depend on the adopted mass
cut. If the ejected amount of $^{56}$Ni ranges between $\sim 0.001$ and $\sim 0.2$ M$_\odot$, the mass
cut remains confined within the region exposed to the complete explosive Si burning and 
hence only the elements produced in this region, i.e., iron, cobalt and nickel, would be
significantly modified by changing the mass cut location (see Limongi, Chieffi \&
Straniero 2001). On the contrary, carbon, nitrogen and oxygen yields, which are synthesized 
in a more external layer of the star, where the modifications induced by the explosive burning are 
less important, are more robust. Obviously, the abundance ratios of any element with respect
to Fe in the ejected matter vary according to the assumed ejected mass of iron. It is 
important to emphasize that the yields used here in the mass range $3\la m/M_\odot\la 40$ are computed 
in a homogeneous and consistent way. On the other hand, the final fate of stars in the mass range 
40-100 M$_\odot$ is highly uncertain because it is not clear if they collapse to a black hole or not. 
Woosley \& Weaver (1995) (see their Figure 7) found that zero metal stars reach the moment of the final 
core collapse with a structure much more compact than that of more metal rich ones. Hence, they found that 
stars more massive than about 40 M$_\odot$ should collapse into a BH under the assumption that the energy 
delivered to the star after the bounce does not significantly depend on the initial mass. In this first study we adopt this
scenario while in a forthcoming one we will investigate how the results change
by assuming that also stars more massive than 40 M$_\odot$ contribute to the chemical enrichment of
the intergalactic medium.

Unfortunately we do not yet have stellar yields computed in the same way from the FRANEC
code for VMOs.  As far as we know, the only complete set of stellar yields
available in the literature for VMOs are those of \citet{obe83}. These authors performed 
full evolutionary calculations of zero-metal stars with initial mass in the range
$100\la m/M_\odot\la 250$ and computed the nucleosynthetic products due to incomplete
oxygen burning in PCSN. One of the most interesting results
from these computations is that few or no iron-peak elements ($50<A<60$) are produced
in the PCSN process. This result contrasts, however, with the more recent calculations of
\citet{arn96} and \citet{heg00}. \citet{arn96} studied the explosion of very massive CO cores 
of different mass and found that important quantities of iron-peak elements can be ejected, depending again on the 
mass-cut position. Computations by \citet{heg00} in a $\sim 70$ M$_\odot$ CO core 
(corresponding to a star of initial mass $\sim 150$ M$_\odot$) produce the same figure. Nevertheless, 
we have compared the stellar yields computed by \citet{obe83} 
with those by \citet{arn96} and \citet{heg00} for the same CO core model and found an 
excellent agreement (differences less than a factor of 2) except, as mentioned previously, for the iron-peak 
elements. Because of the high uncertainty related to the iron yield in VMOs, we decided to adopt that of 
\citet{arn96}. We note that theoretical interpretations of the element ratios 
(in particular the low Ba abundances and the high dispersion in the [Ba/Fe] ratio) found in the most 
metal-poor stars of our galaxy \citep{mac95, sne98}, might indicate that a source other than SNII must exist 
to produce the Fe observed at [Fe/H]$\leq -3$. \citet{was01} attribute this source to the first VMOs 
formed from the Big Bang debris.

Figure 3 shows the yields of the significant elements finally adopted in the stellar mass range
computed by us.

\section{Chemical evolution calculations}

In this section we apply the computed yields and the IMFs proposed for the zero-metal
stars to study the early chemical enrichment of the IGM by means of two different and
simple approaches. In both approaches we consider the entire universe at high redshift
as a one-zone system where the gas exists in a {\it homogeneous} chemical phase. From this
gas with primordial composition, Population III stars form whether isolated, or within
dense clouds with mass $10^5-10^8~M_\odot$ (see $\S 1$). In any case, we
assume that the outflow from the Population III stars is instantly and evenly mixed with the primordial
IGM gas. More realistic models describing the mechanisms from which material processed in stars
formed in primordial clouds is injected and mixed into the IGM, are discussed in \citet{fer00, ma00}. 
This is, however, beyond the scope of the present paper.

First, we consider a single {\it burst} of star formation. In this approach the stellar ejecta is 
incorporated instantaneously into the IGM and there is no further episode of star formation, i.e.:
the ejecta are not incorporated into any second generation of stars. The second case is a simple chemical 
evolution model where the stellar lifetimes are considered explicitly. The model is evolved for $10^8$ yr
in time steps of $10^5$ yr to resolve the small lifetimes of the VMOs, starting at the 
onset of the Population III star formation. This allow us to clearly distinguish between the role played
by VMO, massive stars and IMS respectively, in the pollution of the IGM along the evolving cosmic time. 
Note that a time $\sim 10^8$ yr is roughly the lifetime of a zero-metal star of 3-4 M$_\odot$. 
Although in this second approach several generations of stars are created, and the ejecta of one generation
is mixed into the next generation of stars, the evolution time considered in the model is
short enough to minimize this effect (see below). To give some numbers, placing the very first star 
formation episode at redshift $z\sim 10-20$ in a universe with $\Omega= \Omega_M + \Omega_\Lambda=1$
and H$_o=65$ km/s/Mpc, a time interval $\Delta t\sim 10^8$ yr corresponds to a redshift $z\sim 8.5-15$, respectively. 
For comparison, $z\sim 3$ which is assumed to be the epoch of galaxy formation, corresponds to 
$\Delta t\sim 2 \times 10^9$ yr after the Big Bang. 

\subsection{The burst model}

To compute the chemical enrichment of the IGM due to a single
burst of star formation we proceed as usual. Initially all the
baryons in the universe are in gaseous form with a comoving density
$\rho_b$. During the star burst a fraction $p\equiv\rho_\ast/\rho_b$ (where
$p$ is the star formation {\it efficiency}) goes into stars. Once the generation of stars
have died, a fraction $R$ of their mass is returned into the IGM as processed
gas. This fraction is

\begin{equation}
R=\int_1^{m_{up}}\phi(m) R_{ej}(m) dm
\end{equation}

where $R_{ej}(m)$ is the mass fraction of a star of mass $m$ ejected when the star dies.
This fraction varies depending on the adopted IMF, $\phi(m)$: 0.67, 0.86, 0.82,
0.4 and 0.45 for the Salpeter, YSa, YSb, NUa, NUb IMFs, respectively.
The choice of the IMF also determines the upper mass limit $m_{up}$ (see Figure 1). It is easy
to show that the comoving density of the gas after the burst is given by

\begin{equation}
\rho_{gas}=\rho_b-(1-R)p\rho_b
\end{equation}

On the other hand, if the initial gas density of each isotope is given by $\rho_i^o=\rho_bX_i^o$,
where $X_i^o$ is its primordial abundance in mass fraction, the density of a given isotope
after the burst is

\begin{equation}
\rho_i=\rho_i^o-\rho_\ast X_i^o + \rho_i^{eject}
\end{equation}

where $\rho_i^{eject}$ is the gas density of the specific isotope in the material ejected by
the generation of stars. We thus define the enrichment of the IGM in the isotope $i$ as

\begin{equation}
\Delta X_i={\rho_i\over\rho_{gas}}- X_i^o
\end{equation}

For isotopes with no primordial composition $X_i^o=0$. After some simple algebra from
equations (2) and (3), the chemical enrichment can be written.

\begin{equation}
\Delta X_i={p\over [1-(1-R)p]}(y_i - X_i^o)
\end{equation}

where $y_i$ is the stellar yield defined as the mass fraction of the total mass
returned by a generation of stars in the form of the isotope $i$.

The results from the combined effect of the stellar yields and IMFs for Population III
are summarized in Table 1. The second column gives the total
metallicity $Z$ reached in the material ejected by the star burst (i.e., without 
considering the dilution with the metal-free IGM gas) relative to the solar
metallicity ($Z_\odot=0.0189$). Column three shows the maximum efficient factor 
$p$ allowable for each IMF so as not to surpass the maximum metallicity observed
in the Ly-$\alpha$ forest, $Z\sim 4\times 10^{-3} Z_\odot$. Note that quite low
values of $p$ are allowed for all the IMFs. The most restrictive $p$ values
are those obtained for the NUa \& NUb IMFs because of the high metal pollution
from VMOs ($Z/Z_\odot>1$, see column two)\footnote{As mentioned in $\S 2$, the metallicity
measured in the Ly-$\alpha$ forest systems is uncertain and may well be lower limits. In such
a case, the values of $p$ in Table 1 should be considered also as lower limits. In fact, 
Hernandez \& Ferrara (2001) and Ciardi et al. (2000), estimate slightly larger
values of $p$ than those in Table 1 from different analyses, which would be in agreement
with this figure.}. These low values of $p$ contrast with the current estimate
of the efficient factor in regions of star formation in the Galaxy, $p\la 10^{-2}$, or
in other spiral galaxies \citep{ken98}. However, for any IMF choice in Table 1 the corresponding 
$p$ factor is large enough for the Population III to have played an important role in the reionization 
of the universe before $z\sim 6-7$ (see e.g. Bromm, Kurdritzki \& Loeb 2000).

We can also use the star burst model to estimate the contribution of 
the stellar remnants from Population III to the baryonic matter density. 
It is a simple matter to show that the comoving density of stellar remnants $\rho_{sr}$ is
related to the baryonic density by
 
\begin{equation}
\rho_{sr}=(1-R)p\rho_b 
\end{equation}

Then, if we use $\Omega_{sr}$ to describe the density of stellar remnants in units of the critical density 
($\rho_{crit}=2.77\times 10^{11} h^2$ M$_\odot$ Mpc$^{-3}$), 

\begin{equation}
\Omega_{sr} =(1-R)p \Omega_b 
\end{equation}

where $\Omega_b$ is the baryonic density in units of the critical one. In column four 
of Table 1 we show $\Omega_{sr}$ derived from equation (7), the corresponding $p$ value
compatible with the metallicity limit requirement (see $\S 2$), assuming $\Omega_b=0.019$
and H$_o=65$ km/s/Mpc. The values of $\Omega_{sr}$ are quite low, representing only
a very small fraction of $\Omega_b$ for any IMF choice ($\Omega_{sr}/\Omega_b< 3\times 10^{-3}$).
This can be seen more clearly in Figure 4, where we report the abundance of C versus $\Omega_{sr}$
obtained under different assumptions for the mass function from equation 5 (after a
straightforward transformation). The two vertical lines bound the region compatible with the
baryonic mass budget implied by a Galactic halo interpretation of the LMC MACHO events
\citep{fil98}, whereas the horizontal one indicates the abundance
measured in the most metal-poor Lyman-$\alpha$ forest system. Clearly, for a $\Omega_{sr}\ga
10^{-3}$ model predictions are far beyond of the observational limits for any mass function. 
The low values of $\Omega_{sr}$ required for our analysis to be compatible with the
observational constraints are even more limiting than those imposed
in other studies. For instance, \citet{mad00} impose a limit  from the observed extragalactic 
background light, $\Omega_{sr}/\Omega_b< 5\times 10^{-2}$. The strong limitation on $\Omega_{sr}$ that 
we found here coincides with the conclusion by \citet{fil00} in a similar nucleosynthesis 
analysis, i.e. that the contribution of the stellar remnants from an ancient generation of 
stars to the baryonic dark matter would have been insignificant.
 
Columns 5 to 11 of Table 1 give the IGM abundances relative to the Solar System 
values \citep{and88} for certain elements after the star burst. These abundance 
ratios are computed from equation (5) using the corresponding efficient factor $p$ (column two) for
each IMF entry. Remarkably high [C,N/H] values result from the Salpeter,
YSa and YSb IMFs, due to the large C and N production by zero metal IMS,
which play an important role in these IMFs. In contrast, note the low [N/H] value
obtained in the Nakamura \& Umemura cases. This is because the N yield from VMOs is very
small. However, the [Fe/H] ratio obtained in the Nakamura \& Umemura cases is large if 
Arnett's (1996) iron yield in VMOs is used. Si and Ca production is also
important in these objects in such a way that [Si,Ca/Fe]$\ga 0$ although, [C,N,O,Mg/Fe]$\la 0$. 
Of course, if the iron yield in VMOs from \citet{obe83} is used, the [Fe/H] ratio from equation
5 in the Nakamura \& Umemura cases would be much lower ([Fe/H]$\sim -6$) and, in consequence, extremely high 
[X/Fe] ratios are obtained for any chemical species in Table 1. 

Obviously, the strongest constraints that can be imposed on the nature of the Population III
IMFs come from the ratios between different elements. Unfortunately, the
shortage of abundance measurements in Ly-$\alpha$ forest systems still prevent us from drawing 
limiting conclusions from the results in Table 1. Note in addition, that currently is still
under debate whether the abundances measured in the high redshiftz systems reprensent the
chemical imprint of the Population III stars or not (see $\S 2$). Even if Population III stars
existed, these abundances could have a nonnegligible contribution from non-zero
metallicity stars. Having this in mind, from Table 1 we can see that the best IMF choice 
compatible with the observational constraints discussed in $\S 2$ would be
that of YSa, i.e. an IMF peaked in the high mass range (5-7 M$_\odot$) of IMS, also including 
massive stars with a Salpeter-like slope but without VMOs (see Figure 1). In this case [C/H]$= -2.3$ 
and [Si/H]$= -2.4$ are obtained, in full agreement with the single abundance measurement
of these elements in the Ly-$\alpha$ forest systems. The large [C,N/Fe] ratios obtained 
with the YSa IMF are also compatible with the large C and N overabundances found 
in some extremely metal-poor stars (see references in $\S 2$). Furthermore, small overabundances of 
$\alpha$-elements [O,Mg,Si,Ca/Fe] are also obtained with this IMF, of the order of those derived 
in damped Ly-$\alpha$ systems. In contrast, the [C,N/O]$>0$ ratios obtained
with an IMF that is heavily weighted in the IMS range (cases YSa and YSb mainly) contradict the abundance ratios
([C,N/O]$<0$) found in Population II stars [note however that the O abundance in these stars is still
very controversial, see e.g. \citet{boe99, isr01}]. In this respect it would be very interesting to derive 
O abundances in the Ly-$\alpha$ forest systems and in the extremely metal-poor stars with large
[C,N/Fe] ratios. This might determine whether the IMS were present in the very first generation of stars.

Finally, the last column of Table 1 shows the Li abundance by number (Li/H) expected in the 
material processed after the burst. Since the only evidence of a stellar production of Li comprises the high Li abundances 
measured in some AGB stars \citep{abi91}, we have only considered the Li production by IMS, without taking 
into account the possible production by zero-metal SNII through the $\nu$-mechanism \citep{wos95}. 
From Table 1 it is clear that no significant Li contribution from zero-metal IMS to that produced in 
the Big Bang is expected from any IMF choice. In fact, the Li enrichment $\Delta X_7$ computed from
equation (5) is negative for any value of $p$ assuming $X_7^o\approx 10^{-9}$.

\subsection{The evolutionary model}

We have adopted the \citet{abi95} chemical evolution package, in order to follow the evolution
of some isotopes over time. Current knowledge of star formation history at very
early epochs in the evolution of the universe is very limited \citep{ste99, mad96, hop00}. The best 
estimate available at present indicates a roughly constant star formation activity between $z\sim 1-4$ but no 
data exists for $z>5$. Therefore, as it is usually done, we have assumed a star formation rate 
proportional to the comoving gas density $\psi(t)=\alpha \rho_{gas}^n(t)$, where 
$\alpha= 2$ Gyr$^{-1}$ is the astration parameter and $n=1$. Note that the computed 
abundance ratios [X/Y] are nearly insensitive to the adopted 
star formation rate but they do depend upon the stellar yields and on the IMF adopted. 
In fact, parallel calculations with other $\alpha$ and $n$ values were made to ensure that our
calculations were not dependent upon these parameters. As mentioned previously, we have
followed the evolution of the abundance ratios for $10^8$ yr from the onset of the
primordial star formation. We have not considered the role played by type Ia supernovae
since these objects are usually thought to come from longer lifetime progenitors
\footnote {This is not strictly true in the case of a Red Giant accreting mass on a WD generated by an
intermediate mass progenitors.}. In any case, decisions regarding of the IMF and the law for the star formation
rate have, however, important consequences for the present-day type Ia SN rate (see Canal, Isern 
\& Ruiz-Lapuente 1997). 
The stellar lifetimes have been obtained by means of the same models used to derive the stellar yields. The
computed stellar models  
extend from the pre-main sequence until the thermally pulsing AGB phase for the IMS,
or until the Si-melting in the massive star range. For VMOs
we have extrapolated our numerical lifetime-stellar mass relation to the corresponding mass
range. For example, the lifetimes of stars with initial masses of 4, 25 and 120
M$_\odot$ are 114, 7.8 and 1 Myr, respectively. Due to the very short lifetimes of massive
stars and VMOs, several generations of these stars would form during our adopted evolution
time ($10^8$ yr). Because of the progressive pollution of the IGM, the subsequent
generations of stars would form with a non strictly zero metallicity. Although we also have
yields for massive stars computed with the FRANEC code for different metallicities, as far as
we know there is no mention in the literature of this kind of computation for VMOs [setting aside
the possibility that these objects may form with a non-zero metallicity (see
however Figer et al. 1998)]. Therefore, for the sake of homogeneity in our computations, we
did not consider this effect, i.e. the metallicity dependent yields. We believe nevertheless, 
that this inconsistency in the model
would only have a minimal effect due to the very short time interval considered.

In Table 2 we show the total metallicity reached in the IGM for the
different IMFs at several values of the cosmic time. We consider this metallicity as
typical mean values since at very early epochs the 
pollution of the IGM was very inhomogeneous, probably occurring through the mixing of cloud patches 
already contaminated with metals \citep{arg00}. We emphasize that the
metallicity values in Table 2 are extremely dependent on the star
formation rate history assumed. What is important in Table 2 is the comparison 
between the results obtained with different IMFs at a given cosmic time. As expected from Figure 1, the evolution
of the metallicity obtained in the Salpeter case is between that of the YSa and YSb cases. Note the
very low value of $Z/Z_\odot$ in the YSb case at log t= 6.5. This is due to the steeper slope
of this IMF for the high stellar mass range. In contrast, at log t$\sim 8$ the YSa IMF presents
an important degree of metal enrichment due to the smoother than conventional slope of this IMF for the massive stars
(see Figure 1). The high metallicity value reached $Z\sim 3\times 10^{-2}$ Z$_\odot$ ([Fe/H]$\sim -1.7$) 
at log $t\sim 8$ in the YSa case is similar to that of the low metallicity tail in the metal distribution 
of thick disk stars in our galaxy \citep{nor91}.
For the Nakamura \& Umemura IMF cases, the metallicity in the IGM increases rapidly at very
early times due to the contamination from massive stars and VMOs, but the pollution is not as rapid
at later times as that in the Salpeter and/or Yoshii \& Saio IMFs. This is due to the lesser
importance of IMS which contribute at late in the two former IMFs. 
For the Salpeter, YSb and NUa cases a metallicity similar to that of the most metal-poor
stars in our galaxy is achieved at log t$\sim 8$: [Fe/H]$\approx -2.5$, $-2.7$ and $-3.0$ are
obtained, respectively. Interestingly, these values are of the order of those predicted in the 
prompt-enrichment model for the early galaxy proposed by \citet{was01, was00}.

Figure 5 shows the evolution in time of several abundance ratios for the different IMFs.
We plotted the abundance ratios against the cosmic time instead of [Fe/H] as this 
abundance ratio is very dependent upon the model parameters. The abundance ratios obtained in our model
for a time $\sim 10^8$ yr after the onset of the Population III star formation, might  be interpreted
as those present in the more massive and denser collapsed structures from which the
present-day galaxies would have formed, i.e. the high redshift damped Ly-$\alpha$ systems 
(see $\S 2$). From Figure 5 it can be seen that the evolutions of the abundance ratios computed in the
Salpeter and YSa cases are almost equal. The sole difference appears in the [C/Fe] and
[N/Fe] ratios at late times due to the more important contribution of IMS in the
Salpeter case. For the YSa and Salpeter IMFs, overabundances of [C,N/Fe]$\ga 0$
are obtained at log t$\sim 8$ as well as for the $\alpha$-element ratios [O,Mg,Si,Ca/Fe].
Note the sudden increase of the [N/Fe] ratio at log t$\sim 7.6$ which indicates
the onset of the IMS contribution to the IGM metal enrichment. On the contrary, 
underabundances of [Mg,Si/Fe] are obtained with the YSb IMF. The reason for this
is the steeper slope of this IMF in the mass range of the SNII progenitors from which
these $\alpha$-elements are produced. From the YSb IMF,  however, a sharper increase
of the [C,N,O/Fe] ratios at later times is obtained, which indicates the 
importance of the IMS contribution to CNO in this case (see Figure 1). In fact,
[C,N/Fe]$> 1$ are obtained at log t$>7.6$ for the YSb case

For the Nakamura \& Umemura cases, the evolutions of the abundance ratios are equal (see Figs. 5 and 6)
thus, we have omitted the NUb case from these figures for the sake of clarity. However, the
corresponding evolution of the absolute abundances are rather different, as it can be seen in
Table 2. Again, overabundances of $\alpha$-elements are obtained for any time\footnote{Note that in the
NUa and NUb cases the [Ca/Fe] evolution refers to the combined evolution of 
$\alpha$-elements in the mass range $32\leq A\leq 40$ ($^{32}$S, $^{38}$K and $^{40}$Ca).}.
The evolution of the [C,N/Fe] ratio differs significantly with respect to that obtained
from the Salpeter, YSa and YSb cases. At early times underabundances of C and N are predicted,
in particular that of nitrogen. This is due to the very small yield of $^{14}$N in zero
metal VMOs. However, the large C and N overabundances at late times is
again due to the production by IMS (see Figure 1). From Figure 5 it is obvious that the
evolution of the [C/Fe] and [N/Fe] ratios is dramatically affected by the yields from IMS for
any IMF choice. To emphasize the importance of the zero-metal IMS in the early enrichment of the IGM, 
in Figure 6 we have plotted the evolution of the CNO abundance ratios. Extreme 
[C/O] and [N/O] ratios result when an IMF that is heavily weighted around 2-4 M$_\odot$ is used 
(YSb, NUa and NUb cases). Note also the very low [N/O,C] ratios predicted at early times when the presence of 
VMOs is considered in the IMF.

How can the evolution of the abundance ratios in Figures 5 and 6 be related to observations? 
If we consider only the abundance ratios obtained at log t$\sim 8$, the abundance ratio
[Si/C]$\sim +0.2$ measured toward the Ly-$\alpha$ forest systems \citep{son97} would be
in better agreement with a Salpeter IMF. A steeper slope in the mass range $10\la m/M_\odot\la 40$, as 
in the YSb case, does not seem adequate. The presence of VMOs in the primordial IMF is also
favored by this [Si/C] ratio provided that an important C production from IMS is excluded. On the contrary, 
on the basis of the present yield calculations for zero-metal stars, the only way to obtain large
[C,N/Fe]$>1$ ratios is to accept an IMF peaking in the IMS range. An IMF of such a type
could explain the high C and N abundance ratios found in a considerable number of extremely
metal-poor stars. This, however, would imply high [C,N/O] ratios. Unfortunately, there exist
no oxygen abundance determinations in these very metal-poor stars to test this hypothesis.
Note that in Population II halo stars with [Fe/H]$<-1.5$ the observed ratios are [C,N/O]$\sim -0.5$
\citep{tom92}. Although
the O abundances are also uncertain in these stars, these low [C,N/O] ratios would be against
an IMF heavily weighted of intermediate mass stars. In this respect, our conclusions are similar to those reached by
\citet{gib97} using $Z\sim 10^{-4}$ Z$_\odot$ stellar yields for IMS. 
On the other hand, the [O/Fe] ratios derived in the most metal-poor stars in our Galaxy certainly 
put interesting constraints on the very early chemical evolution. With the yields for zero-metal stars used here, 
we did not find ratios [O/Fe]$\ga 1$ as derived in several recent studies \citep{abi89, isr98, boe99, isr01, 
mis00} for any IMF. However, taking into account the uncertainty in the iron yield from massive and
very massive stars (see $\S 4$), it is indeed possible to obtain [O/Fe]$>1$ or higher at very early times
by decreasing the iron yield from these stars with respect to that adopted here, by only a factor of four. The present 
uncertainty on the iron yield from these stars, allows such a variation.

\section {Summary and conclusions} 
We have studied the chemical constraints imposed on the existence of a pregalactic
generation of stars (Population III) by the recent abundance determinations in
high redshift systems and in extremely metal-poor stars in our Galaxy. Using yields computed
from stellar models ($3\la m/M_\odot\la 200$) of strictly zero metal content, we have 
shown that a chemical enrichment to the level observed in the high redshift
intergalactic medium can be easily obtained from a stellar pregalactic nucleosynthesis. We merely have
to postulate (see Table 1) that a small fraction ($p<10^{-2}$) of primordial matter has
participated in the Population III stars. As a consequence, this low pregalactic star
formation efficiency strongly limits the contribution of the stellar remnants from Population
III to the baryonic matter. We found $\Omega_{sr}< 10^{-3}\Omega_b$ for all of the IMFs tested. 
However, the present scarcity of abundance data in the high redshift universe (which are of
controversial interpretation) does not yet permit us to strongly limit the IMFs currently proposed
for Population III: basically, an IMF peaking in the intermediate stellar mass range 
($3\la m/M_\odot\la 8$) or an IMF including very massive objects ($m\ga 100$ M$_\odot$). Nevertheless,
at the present state of the art on stellar nucleosynthesis models, the very 
large C and N enhancement ([C,N/Fe]$>1$) found in a significant fraction of the extremely metal-poor stars 
in our Galaxy favors a Population III IMF peaked at intermediate mass stars.  However, as it has
also been shown by previous studies \citep{gib97, fil00}, this inevitably leads to a pollution of the halo 
ISM at the levels of [C,N/O]$>0.5$, which seems rather difficult considering the observed [C,N/O] abundance
pattern in Population II halo dwarfs. 

Without doubt, further abundance studies in high redshift systems, in particular in Lyman$-\alpha$ forest
systems with decreasing hydrogen column density, and in extremely metal-poor stars will soon resolve the
question of the existence of Population III and the nature of its mass function.

\acknowledgments
This work has been partially supported by the grants PB96-1428, AYA2000-1574 and the Italy-Spain 
agreement HI1998-0095, the PNIE, the CIRIT and the italian grant COFIN 2000.

\clearpage

\begin{deluxetable}{cccc}
\footnotesize
\tablenum{2}
\tablecaption{Evolution in time of IGM metallicity ($Z/Z_\odot$) for the different IMFs.}
\tablewidth{230pt}
\tablehead{
\colhead{log t}& \colhead{6.5} & \colhead{7.3}    & \colhead{8}
}
\startdata

Salpeter&$4.2\cdot 10^{-7}  $ & $2.1\cdot 10^{-4}$ & $3.1\cdot 10^{-3}$\\
YSa     &$1.6\cdot 10^{-7}  $ & $2.1\cdot 10^{-3}$ & $3.1\cdot 10^{-2}$\\
YSb     &$1.6\cdot 10^{-10} $ & $4.2\cdot 10^{-5}$ & $2.4\cdot 10^{-3}$\\
NUa     &$2.2\cdot 10^{-5}  $ & $1.1\cdot 10^{-4}$ & $1.0\cdot 10^{-3}$\\
NUb     &$3.2\cdot 10^{-7}  $ & $1.6\cdot 10^{-6}$ & $1.7\cdot 10^{-5}$\\

\enddata
\end{deluxetable}
\clearpage

\figcaption[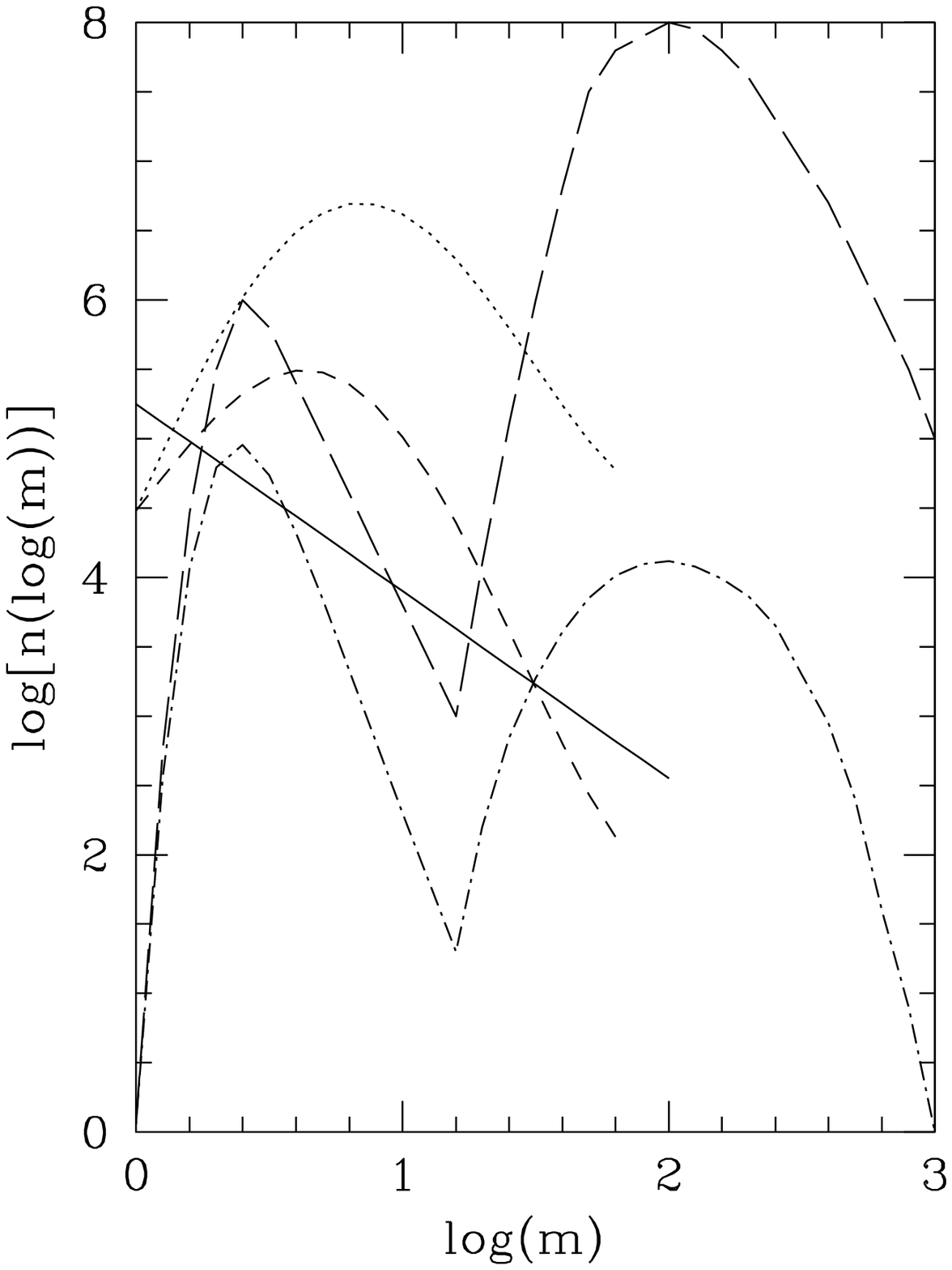]{Comparison of the IMFs used in this work over the
range $1\leq m/M_\odot\leq 10^3$. Solid line: Salpeter mass spectrum. Dotted line:
YSa. Short dashed line: YSb. Long dashed line: NUa. Dashed-dotted line: NUb.\label{fig1}} 

\figcaption[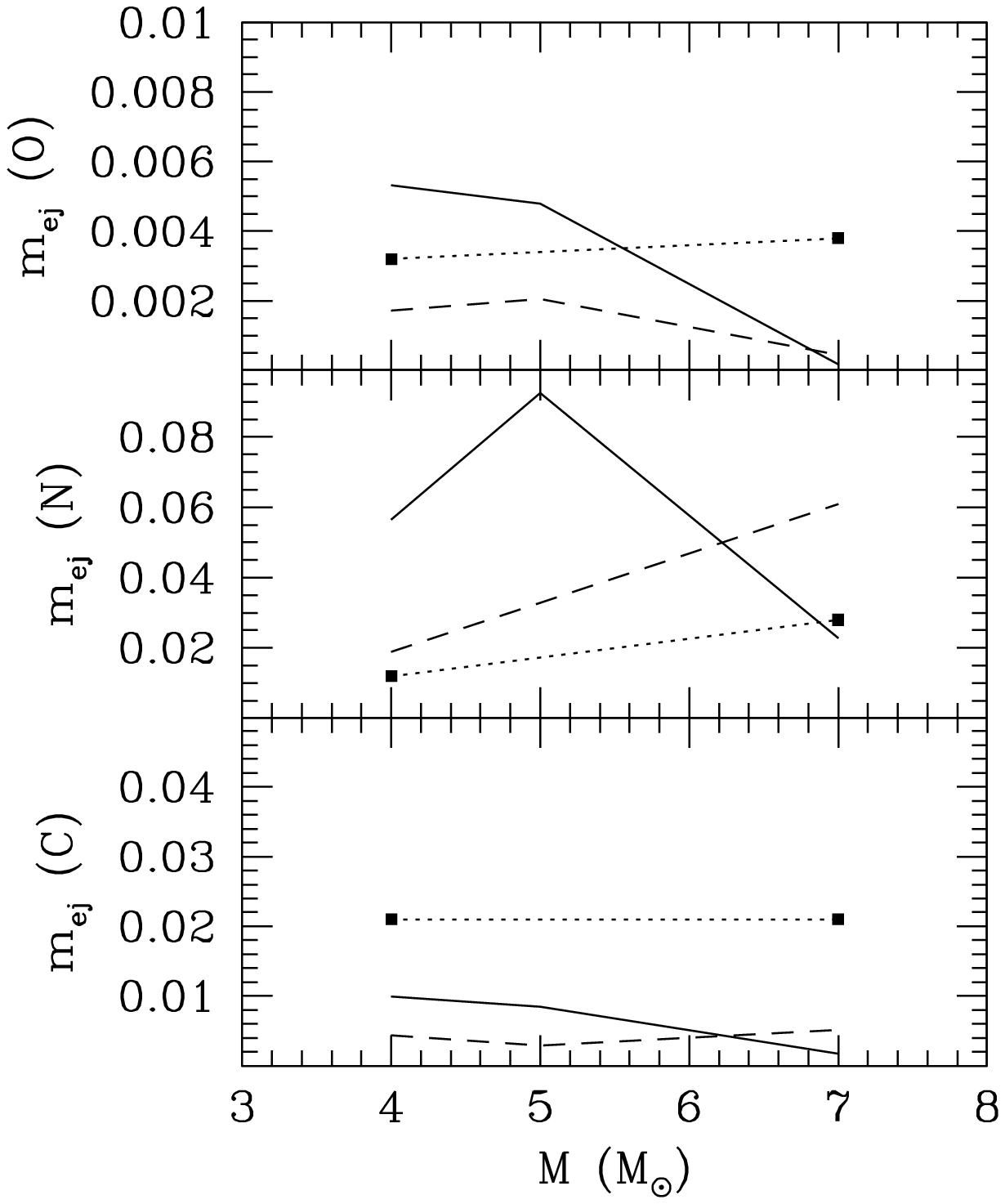]{Yields (i.e total ejected mass in solar masses) from IMS predicted by
van der Hoek \& Gronewegen (1997), for $Z=10^{-3}$ and $\eta=1$ (solid lines) and
$\eta=4$ (dashed lines), and those computed by Chieffi et al. (2001) for
$Z=0$ and $\eta=1$ (dotted lines and squares).\label{fig2}} 

\figcaption[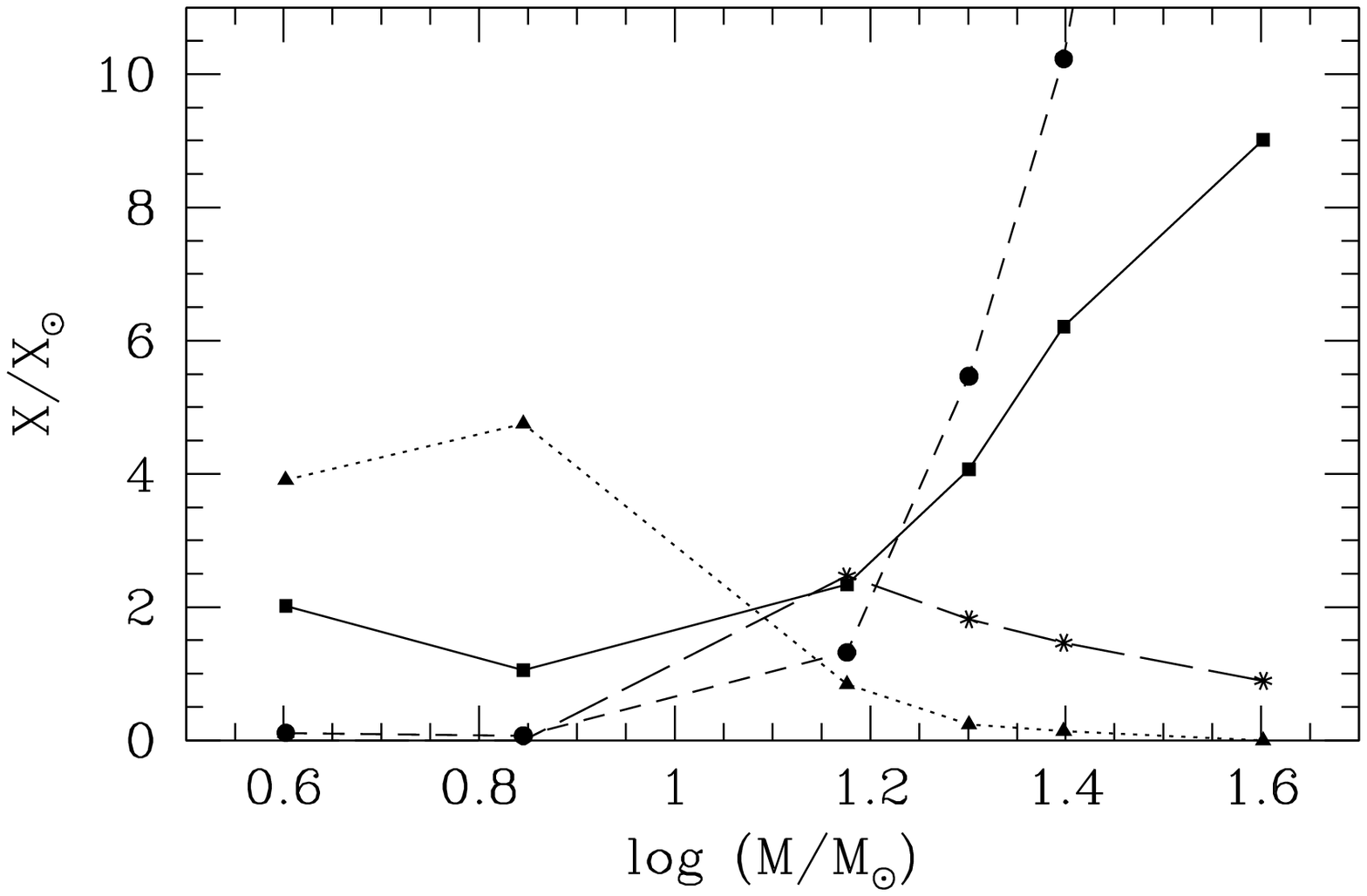]{Adopted production factors as a function of the stellar mass for
certain elements: C (filled squares), N (filled triangles), O (filled circles) and
Fe (stars). Note that the oxygen production for a 40 M$_\odot$ star is well above
the figure limits: $X_O/X_\odot\sim 27$.\label{fig3}}

\figcaption[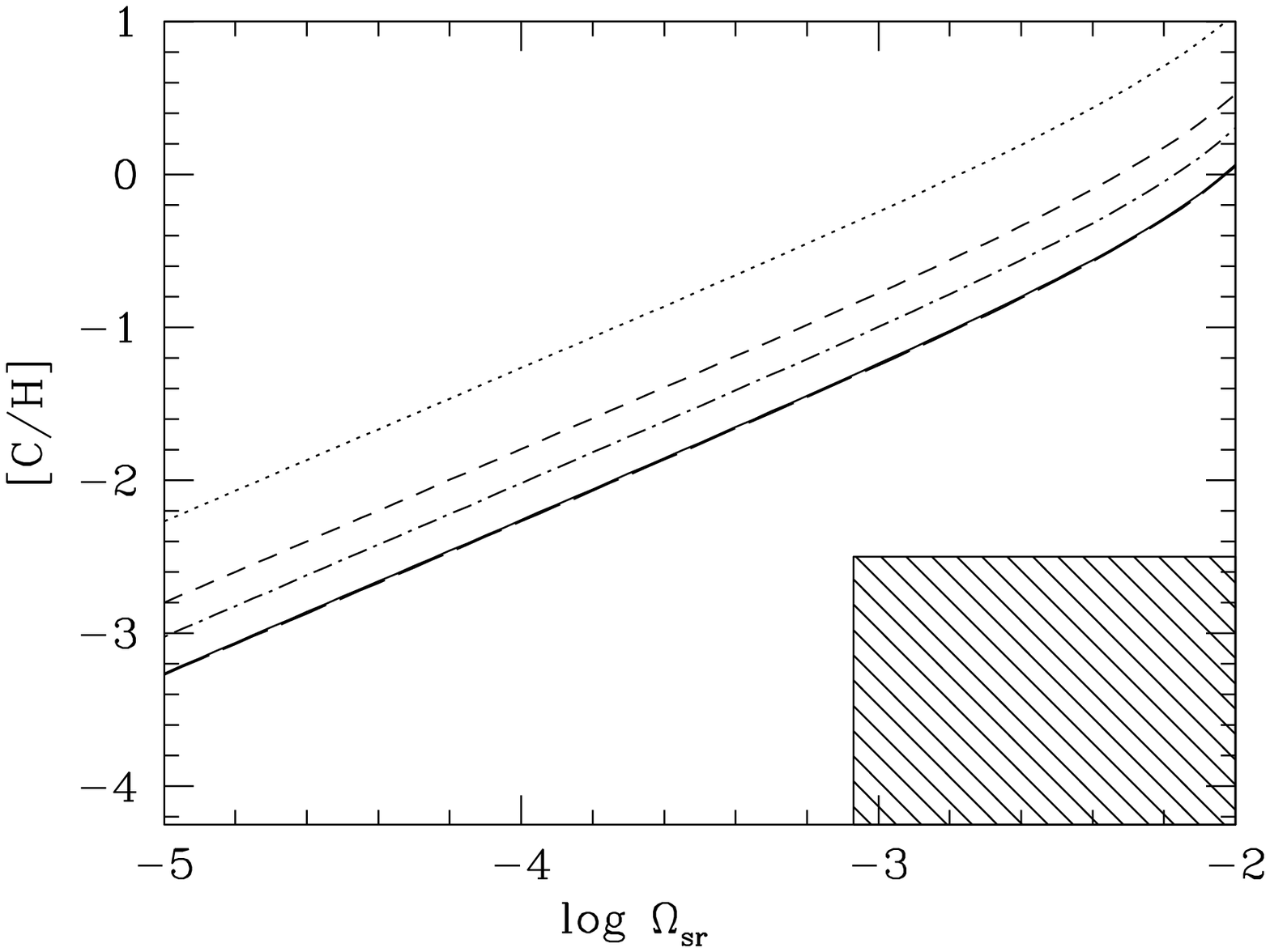]{Carbon production from Population III stars as a function
of the remnant density adopting $\Omega_b=0.019$ and H$_o=65$ km/s/Mpc. The various lines represent the results
obtained with the IMFs reported in Figure 1. The shaded box define the region in which [C/H]$\la -2.5$
(see $\S 2$) and $8\times 10^{-4}\la \Omega_{sr}\la 10^{-2}$ according to
\citet{fil98}. Key for the curves as in Figure 1. Note that the Salpeter and NUa curves
nearly coincide.\label{fig4}}

\figcaption[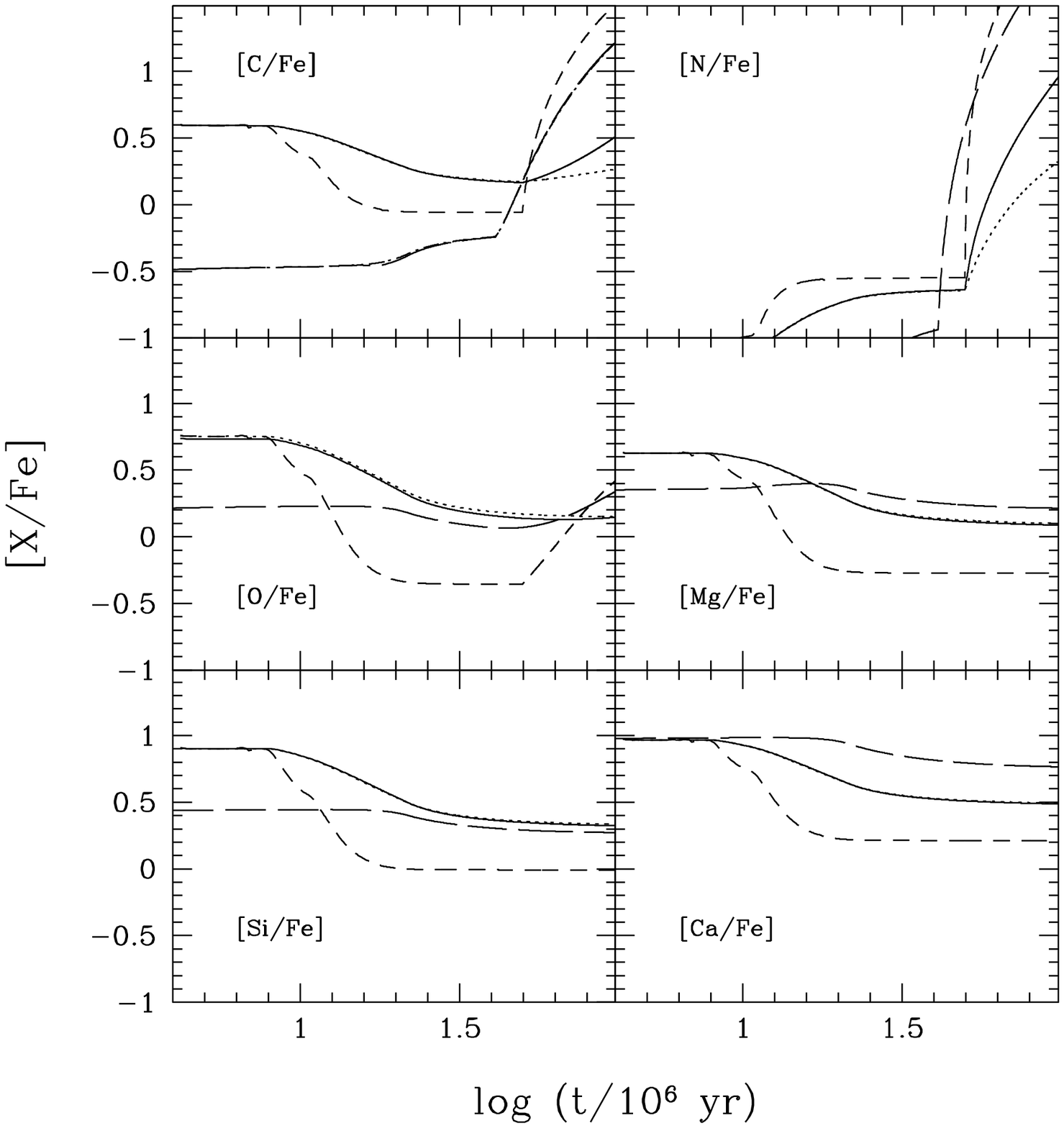]{Evolution of the different abundance ratios in time for the different 
IMF choices. Key for the curves as in Figure 1.\label{fig5}}

\figcaption[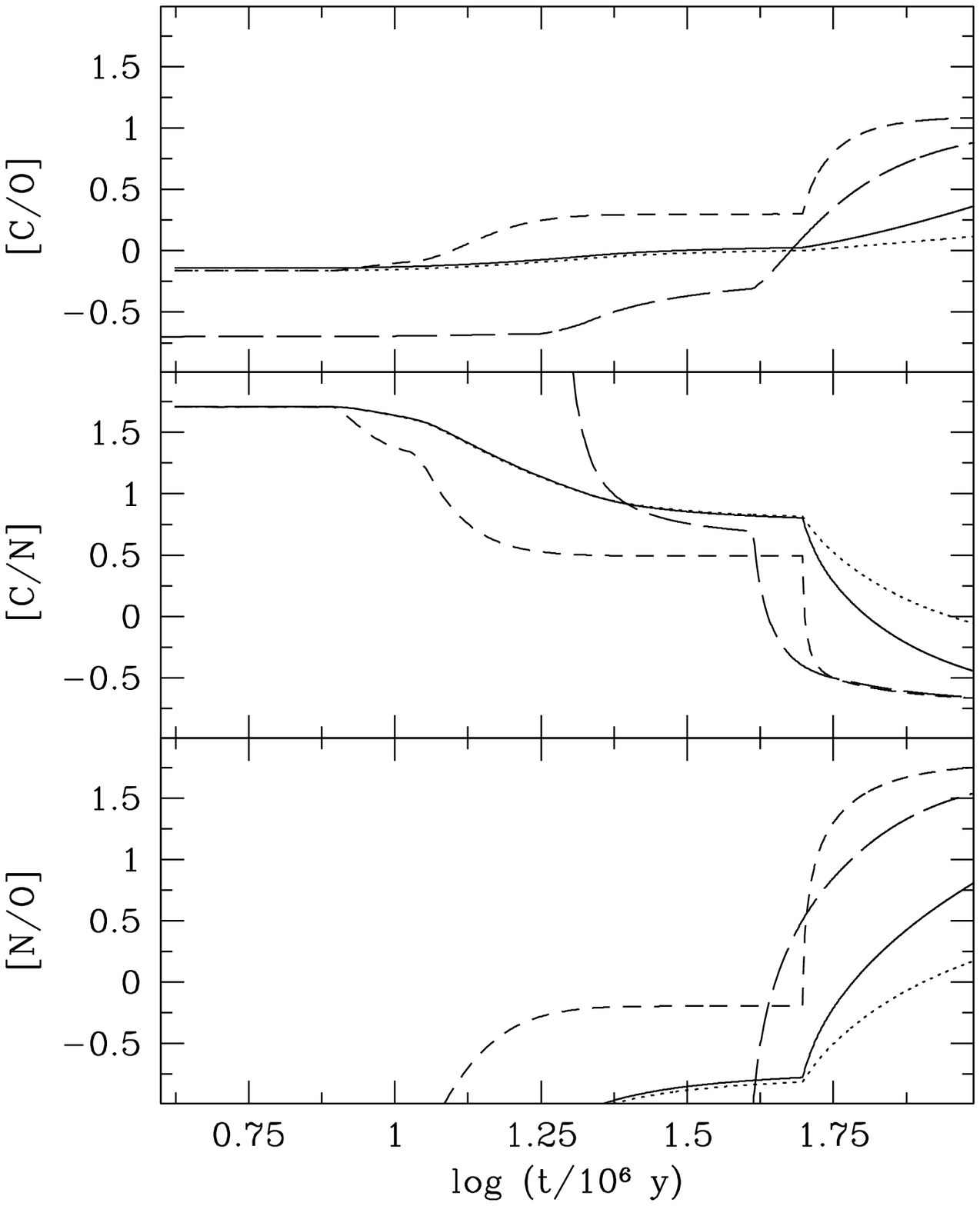]{Evolution of the IGM [C/O], [C/N] and [N/O] ratios for the model described in 
$\S 5$. Key for the curves as in Figure 1.\label{fig6}} 

\clearpage 
\plotone{fig1.eps}
\clearpage
\plotone{fig2.eps}
\clearpage
\plotone{fig3.eps}
\clearpage
\plotone{fig4.eps}
\clearpage
\plotone{fig5.eps}
\clearpage
\plotone{fig6.eps} 


\begin{thebibliography}{}
\bibitem[Abia \& Rebolo (1989)]{abi89} Abia, C., \& Rebolo, R. 1989, \apj, 347, 186
\bibitem[Abia et al. (1991)]{abi91} Abia, C., Boffin, H.M.J., Isern, J., Rebolo, R. 1991, \aap, 245, L1
\bibitem[Abia, Isern \& Canal (1995)]{abi95} Abia, C., Isern, J. \& Canal, R. 1995, \aap, 298, 465
\bibitem[Alcock, C. et al. (2000)]{alc00} Alcock, C. et al. (The MACHO group) 2000, \apj, 542, 281
\bibitem[Anders \& Grevesse (1988)]{and88} Anders, E. \& Grevesse, N. 1988, Geochim. Cosmochim. Acta,
53, 197
\bibitem[Argast et al. (2000)]{arg00} Argast, D., Samland, M., Gerhard, O.E., \& Thielemann, F.-K. 2000,
\aap, 356, 378
\bibitem[Arnett (1996)]{arn96} Arnett, D., 1996, Supernovae and Nucleosynthesis, Princeton University
Press, p. 375
\bibitem[Beers, Preston \& Shectman (1985)]{bee85} Beers, T.C., Preston, G.W., \& Shectman, S.A. 1985,
\aj, 90, 2089
\bibitem[Beers, Preston \& Shectman (1992)]{bee92} Beers, T.C., Preston, G.W., \& Shectman, S.A. 1992,
\aj, 103, 1987
\bibitem[Boesgaard et al. (1999)]{boe99} Boesgaard, A.M., King, J.R., Deliyannis, C.P. \& Vegt, S.S. 1999,
\aj, 117, 429
\bibitem[Bonifacio et al. (1998)]{bon98} Bonifacio, P., Molaro, P., Beers, T.C. \& Vladilo, G. 1998,
\aap, 332, 672
\bibitem[Borksemberg et al. (1998)]{bor98} Borksemberg, A., Sargent, W.L.W., \& Rauch, M. 1998, in The
Birth of Galaxies, ed. B. Guiderdoni, F.R. Bouchet, Trinh X. Thuan and Tranh Thanh Van (Paris:
Editions Frontieres, in press.
\bibitem[Bromm, Kudritzki \& Loeb (2000)]{bro00} Bromm, V., Kudritzki, R.P., \& Loeb, A. 2000, astro-ph/0007248 
\bibitem[Cameron \& Fowler (1971)]{cam71} Cameron, A.G.W., \& Fowler W.A. 1971, \apj, 164, 111
\bibitem[Canal, Isern \& Ruiz-Lapuente (1997)]{can97} Canal, R., Isern, J., \& Ruiz-Lapuente, P. 1997, \apj,
488, L35
\bibitem[Carberlg (1981)]{car81} Carlberg, R.G. 1981, \mnras, 197, 1021
\bibitem[Cayrel (1986)]{cay86} Cayrel, R. 1986, \aap, 168, 197
\bibitem[Ciardi et al. (2000)]{cia00} Ciardi, B, Ferrara, A., Governato, F., \& Jenkins, A. 2000, \mnras, 314, 611
\bibitem[Chieffi, Limongi \& Straniero (1998)]{chi98} Chieffi, A., Limongi, M., \& Straniero 1998,
\apj, 737, 762
\bibitem[Chieffi et al. (2001)]{chi01} Chieffi, A., Dom\'\i nguez, I., Straniero, O., \& Limongi, M. 2001,
\apj, (in press)
\bibitem[Cowie et al. (1995)]{cow95} Cowie, L.L., Songaila, A., Kim, T.S., \& Hu, E. 1995, \aj, 109, 1522
\bibitem[Cowie \& Songaila (1998)]{cow98} Cowie, L.L. \& Songaila, A. 1998, Nature, 394, 248
\bibitem[Depagne et al. (2000)]{dep00} Depagne, E., Hill, V., Christlib, N., \& Primas, F., 2000, \aap (in press)
\bibitem[Ellison et al. (2000)]{ell00} Ellison, S.L., Songaila, A., Schaye, J., \& Pettini M. 2000,
\aj, 120, 1167
\bibitem[Fan et al. (2000)]{fan00} Fan, X. et al. 2000, astro-ph/0005414
\bibitem[Ferrara, Pettini \& Shchekinov (2000)]{fer00} Ferrara, A., Pettini, M., \& Shchekinov, Y. 2000,
astro-ph/0004349
\bibitem[Fields, Freese, \& Graff (1998)]{fil98} Fields, B.D., Freese, K., \& Graff, D. 1998, NewA, 3, 347
\bibitem[Fields, Freese \& Graft (2000)]{fil00} Fields, B.D., Freese, K., \& Graft, D.S. 2000, \apj, 534, 265
\bibitem[Figer et al. (1998)]{fig98} Figer, D.F., Najarro, F., Morris, M., McLean, I.S., Geballe, T.R.,
Ghez, A.M., \& Langer, N. 1998, \apj, 506, 384
\bibitem[Fujimoto, Ikeda \& Iben (2000)]{fuj00} Fujimoto, M.Y., Ikeda, Y., \& Iben, I.Jr. 2000, \apj, 529, L25
\bibitem[Gibson \& Mould (1997)]{gib97} Gibson, B.K., \& Mould, J.R. 1997, \apj, 482, 98
\bibitem[Gnedin (2000)]{gne00} Gnedin, N.Y. 2000, \apj, 542, 535
\bibitem[Heger et al. (2000)]{heg00} Heger, A., Baraffe, I., Fiyer, C.L., \& Woosley, S.E. 2000, Nucl.
Phys. (in press)
\bibitem[Hernandez \& Ferrara (2001)]{her01} Hernandez, X., \& Ferrara, A. 2001, \mnras, in press
\bibitem[Hill et al. (2000)]{hil00} Hill, V., Barbuy, B., Spite, M., Spite, F., Cayrel, R., Plez, B., Beers, T.C.,
Nordstr\"om, B., \& Nissen, P.E. 2000, \aap, 353, 557
\bibitem[Hopkins, Connolly \& Szalay (2000)]{hop00} Hopkins, A.M., Connolly, A.J., \& Szalay, A.S. 2000, \aj, 120, 2843
\bibitem[Isern et al. (1998)]{ise98} Isern, J., Garc\'\i a-Berro, E., Hernanz, M., Motchkovich, R., \&
Torres, S. 1998, \apj, 503, 239
\bibitem[Isern et al. (1999)]{ise99} Isern, J., Hernanz, M., Garc\'\i a-Berro, E., \& Motchkovich, R. 1999,
in XI European Workshop on White Dwarfs. S.E. Solhein and E.G. Meistas eds. PASP Publishers, p. 408
\bibitem[Israelian, Garc\'\i a-L\'opez \& Rebolo (1998)]{isr98} Israelian, G., Garc\'\i a-L\'opez, R., \&
Rebolo, R. 1998, \apj, 507, 805
\bibitem[Israelian et al. (2001)]{isr01} Israelian, G., Rebolo, R., Garc\'\i a-L\'opez, R., Bonifacio, P., 
Molaro, P., Basri, G., \& Shchukina, N. 2001, \apj, (in press)
\bibitem[Kennicutt (1998)]{ken98} Kennicutt, R. C. Jr. 1998, \araa, 36, 198
\bibitem[Laserre et al. (2000)]{las00} Laserre, T., et al. (The EROS group) 2000, \aap, 355, L39
\bibitem[Limongi, Straniero \& Chieffi (2000)]{lim00} Limongi, M., Straniero, O., \& Chieffi, A. 2000,
\apjs, 129, 625
\bibitem[Limongi, Chieffi \& Straniero (2001)]{lim01} Limongi, M., Chieffi, S., \& Straniero, O. 2001, \apj, in preparation
\bibitem[Lu et al. (1998)]{lu98} Lu, L., Sargent, W.L.W., Barlow, T.A., \& Rauch, M. 1998, \aj, (astr0-ph/9802189) 
\bibitem[Madau et al. (1996)]{mad96} Madau, P., Ferguson, H.C., Dickinson, M., Giavalisco, M., Steidel, C.C., \&
Fruchter, A. 1996, \mnras, 283, 1388.
\bibitem[Madau \& Pozzetti (2000)]{mad00} Madau, P., \& Pozzetti, L. 2000, \mnras, 312, L9
\bibitem[Madau, Ferrara \& Rees (2000)]{ma00} Madau, P., Ferrara, A., \& Rees, M.J. 2000, astro-ph/0010158
\bibitem[Marigo, Bressan \& Chiosi (1998)]{mar98} Marigo, P., Bressan, A., \& Chiosi, C. 1998, \aap, 331, 564
\bibitem[Matsuda, Sato \& Takeda (1969)]{mat69} Matsuda, T., Sato, H., \& Takeda, H. 1969, Prog. Theor.
Phys., 42, 219
\bibitem[McWilliam et al. (1995)]{mac95} MacWillian, A., Preston, G.W., Sneden, C., \& Searle L. 1995, \aj,
105, 2757
\bibitem[M\'endez \& Minniti (2000)]{men00} M\'endez, R.A., \& Minniti, D. 2000, \apj, 529, 911
\bibitem[Miralda-Escud\'e \& Rees (1998)]{mir98} Miralda-Escud\'e, J., \& Ress, M.J. 1998, \apj, 497, 21
\bibitem[Miralda-Escud\'e (2000)]{mir00} Miralda-Escud\'e, J. 2000, in The First Stars, Weiss A., Abel, T.,
Hill, V. (eds.), Springer, Berlin, p. 242
\bibitem[Mishenina et al. (2000)]{mis00} Mishenina, V.G., Korotin, S.A., Klochkova, V.G., \& Panchuk, V.E. 2000
\aap, 353, 978
\bibitem[Nakamura \& Umemura (2000)]{nak00} Nakamura, F., \& Umemura, M. 2000, \apj, 515, 239
\bibitem[Norris \& Ryan (1991)]{nor91} Norris, J.E., \& Ryan, S.G. 1991, \apj, 380, 403
\bibitem[Norris, Beers \& Ryan (2000)]{nor00} Norris, J.E., Beers, T.C., \& Ryan, S.G. 2000, \apj, 540, 456
\bibitem[Ober, El Eid \& Fricke (1983)]{obe83} Ober, W.W., El Eid, M.F., \& Fricke, K.J. 1983, \aap, 119, 61
\bibitem[Pagel (1997)]{pag97} Pagel, B.E.J. 1997, Nucleosynthesis and Chemical Evolution of
Galaxies. Cambridge University Press.
\bibitem[Pettini (1999)]{pet99} Pettini M. 1999, in Chemical Evolution from Zero to High
Redshift. (Berlin: Springer). J. Walsh and M. Rosa (eds.). Lectures Notes in Pyhsics (in press).
\bibitem[Prochaska \& Wolfe (1999)]{pro99} Prochaska, J.X., \& Wolfe, A.M. 1999, \apjs, 121, 369
\bibitem[Prochaska \& Wolfe (2000)]{pro00} Prochaska, J.X., \& Wolfe, A.M. 2000, \apj, 533, L5
\bibitem[Prochaska, Gawiser, \& Wolfe (2001)]{pro01} Prochaska, J.X., Gawiser, E., \& Wolfe, M.A. 2001, \apj, (astro-ph/0101029)
\bibitem[Rossi, Beers \& Sneden (1999)]{ros99} Rossi, S., Beers, T.C., \& Sneden, C. 1999, The Third
Stromlo Symposium: The Galactic Halo. ASP Conference Series, Vol. 165, p. 268. B.K. Gibson, T.S.
Axelrod T.S. and Potman M.E. (eds.)
\bibitem[Sackmann \& Boothroyd (1992)]{sac92} Sackamann, I.J., \& Boothroyd, A. 1992, \apj, 392, L71
\bibitem[Salpeter (1955)]{sal55} Salpeter, E.E. 1955, \apj, 121, 161
\bibitem[Sargent et al. (1980)]{sar80} Sargent, W.L.W., Young, P.J., Borksenberg, A., Tytler, D. 1980,
\apjs, 42, 41
\bibitem[Schaerer, Izotov \& Charbonnel (2000)]{sc00} Schaerer, D., Izotov, Y.I., \& Charbonnel, C.
2000, in The Evolution of Galaxies, Kluwer Academic Publishers, J.M. Vilchez and S. Stasinska (eds)
(in press)
\bibitem[Silk (1983)]{sil83} Silk, J. 1983, \mnras, 205, 705
\bibitem[Sneden et al. (1996)]{sne96} Sneden, C., McWillian, A., Preston, W., Cowan, J.J., Burris, D.L.,
\& Arnosky, B.J. 1996, \apj, 467, 819
\bibitem[Sneden et al. (1998)]{sne98} Sneden, C., Cowan, J.J., Burris, D.L., \& Truran, J. 1998, \apj, 496, 235
\bibitem[Sneden et al. (2000)]{sne00} Sneden, C., Cowan, J.J., Ivans, I.I., Fuller, G.M., Burles, S., 
Beers, T.C., \& Lowler, J.E. 2000, \apj, 533, L139
\bibitem[Songaila \& Cowie (1996)]{son96} Songaila, A., \& Cowie, L.L. 1996, \aj, 112, 335
\bibitem[Songaila (1997)]{son97} Songaila, A. 1997, \apj, 490, L1
\bibitem[Spite et al. (2000)]{spi00} Spite, M., Depagne, E., Nordstr\"om, B., Hill, V., Cayrel, R., Spite,
F., \& Beers, T.C. 2000, \aap, 360, 1077
\bibitem[Steidel et al. (1999)]{ste99} Steidel, C.C., Adelberger, K.L., Giavalisco, M., Dickinson, M., \&
Pettini, M. 1999, \apj, 519, 1
\bibitem[Tomkin et al. (1992)]{tom92} Tomkin, J., Lemke, M., Lambert, D., \& Sneden, C. 1992, \aj, 104, 1568
\bibitem[Tytler et al. (1995)]{tyl95} Tytler, D., Fan, X.-M., Burles, S., Cottrell, L., Davis, C., Kirkman, D.,
\& Zuo, L. 1995, in QSO Apsorption Lines, ed. G. Meylan (Berli: Springer), 289
\bibitem[Uehara, Susa \& Nishi (1996)]{uea96} Uehara, H., Susa, H., \& Nishi, R. 1996, \apj, 473, L95
\bibitem[Van den Hoek \& Groenewegen (1997)]{van97} Van den Hoek, L.B., \& Groenewegen, M.A.T. 1997, \aaps, 123, 305 
\bibitem[Wasserburg \& Qian (2000b)]{was00} Wasserburg, G.J., \& Qian, Y.-Z. 2000b, \apj, 538, L99
\bibitem[Wasserburg \& Qian (2000a)]{was01} Wasserburg, G.J., \& Qian, Y.-Z. 2000a, \apj, 529, L21
\bibitem[Weaver \& Woosley (1980)]{wos80} Weaver, T.A., \& Woosley, S. E. 1980, Ann. NY. Acad.
\bibitem[Wolfe (1995)]{wol95} Wolfe, A.M. 1995, in The Physics of Insterstellar Medium and Intergalactic
Medium. ASP Conference Series, Vol 80. p.478. A. Ferrara, C.F. McKee, C. Heiles, P.R. Shapiro (eds)
\bibitem[Woosley \& Weaver (1995)]{wos95} Woosley ,S.E., \& Weaver, T.A. 1995, \apjs, 101, 181
\bibitem[Yoshii \& Saio (1986)]{yos86} Yoshii, Y., \& Saio, H. 1986, \apj, 301, 569

\end{thebibliography}
\end{document}